\pdfoutput=1

\documentclass[sigconf,nonacm]{acmart}

\usepackage{subfiles}

\usepackage{amsmath}

\usepackage{bm}
\usepackage{siunitx}

\usepackage{algorithm}
\usepackage{algorithmic}

\usepackage{booktabs}
\usepackage{float}

\usepackage{subcaption}

\usepackage{url}

\newcommand{\aisaq}{AiSAQ }
\newcommand{\aisaqdesc}{All-in-Storage ANNS with Product Quantization }
\newcommand{\imagebasepath}{images}
\title{\aisaq: \aisaqdesc 
       for DRAM-free Information Retrieval}

\author{Kento Tatsuno}
\email{kento1.tatsuno@kioxia.com}
\orcid{0009-0006-6180-2340}
\affiliation{
    \institution{Kioxia Corporation}
    \country{Japan}
}
\authornote{Both authors contributed equally to the paper.}

\author{Daisuke Miyashita}
\email{daisuke1.miyashita@kioxia.com}
\orcid{0000-0003-2108-3397}
\affiliation{
    \institution{Kioxia Corporation}
    \country{Japan}
}
\authornotemark[1]

\author{Taiga Ikeda}
\email{taiga1.ikeda@kioxia.com}
\orcid{0009-0006-5066-5387}
\affiliation{
    \institution{Kioxia Corporation}
    \country{Japan}
}

\author{Kiyoshi Ishiyama}
\email{kiyoshi1.ishiyama@kioxia.com}
\orcid{0009-0004-9891-8430}
\affiliation{
    \institution{Kioxia Corporation}
    \country{Japan}
}

\author{Kazunari Sumiyoshi}
\email{kazunari.sumiyoshi@kioxia.com}
\orcid{0009-0001-5893-0345}
\affiliation{
    \institution{Kioxia Corporation}
    \country{Japan}
}

\author{Jun Deguchi}
\email{jun.deguchi@kioxia.com}
\orcid{0000-0002-3414-5537}
\affiliation{
    \institution{Kioxia Corporation}
    \country{Japan}
}

\date{April 9, 2024}

\setcopyright{cc}
\setcctype{by}

\begin{document}
    
    \begin{abstract}
        Graph-based approximate nearest neighbor search (ANNS) algorithms work effectively against large-scale vector retrieval. Among such methods, DiskANN achieves good recall-speed tradeoffs using both DRAM and storage. DiskANN adopts product quantization (PQ) to reduce memory usage, which is still proportional to the scale of datasets. In this paper, we propose \aisaqdesc (\aisaq), which offloads compressed vectors to the SSD index. Our method achieves $\sim$10 MB memory usage in query search with billion-scale datasets without critical latency degradation. \aisaq also reduces the index load time for query search preparation, which enables fast switch between muitiple billion-scale indices.This method can be applied to retrievers of retrieval-augmented generation (RAG) and be scaled out with multiple-server systems for emerging datasets. Our DiskANN-based implementation\footnote{\url{https://github.com/KioxiaAmerica/aisaq-diskann}} is available on GitHub.


    \end{abstract}

    \maketitle

    \section{Introduction}
        In tasks of retrieving specific data that users are interested in from large-scale datasets such as images, music, and documents, Approximate Nearest Neighbor Search (ANNS), which gets close vectors to query vectors using pre-built indices constructed from the datasets, is commonly utilized. Usually, the index data is stored in fast memories (RAM) to shorten the search latency. However, the DRAM costs too much to store large datasets, especially those that reach a billion-scale, and slower but cheaper storages such as SSDs also need to be employed. 

        In the recent trend of Large Language Models (LLM), possible implementations of Retrieval-Augmented Generation (RAG)\cite{lewis2020retrieval} use vector datasets as their external knowledge sources to generate more accurate answers. In the original paper, Dense Passage Retrieval\cite{dpr2020} is used as a ``retriever'' for seaching the external sources. 
        That retriever uses HNSW\cite{malkov2018hnsw} for searching vector data of external sources, and it can be replaced with other suitable ANNS algorithms depending on the demand of applications.
        In some situations that a RAG chain requires multiple sources, the retriever need to ``switch'' between indices. This procedure requires the index data of all sources to be retained in RAM during the service time or loaded from storage every time of a request.
    
        DiskANN\cite{jayaram2019diskann} is the de-facto standard of graph-based ANNS methods exploiting storages. This is adopted for the baseline of Big-ANN competion track of NeurIPS\cite{bigannbench2023} and is employed in vector databese services such as Weaviate\cite{weaviate-github} and Zilliz \cite{milvus2021wang}. DiskANN claims to acheive query search of billion-scale datasets with high recall and small memory usage. It reduces memory usage by compressing node vectors using Product Quantization(PQ) \cite{jegou2010pq} and keeps high search recall by re-ranking all nodes in the search path by their full-precison vectors loaded from the storage. However, since DiskANN retains PQ-compressed vectors of all nodes in RAM, the memory usage is proportional to the scale of the datasets. 
        Additionally, increasing compression ratio reduces both of memory usage and recall, which leads to the trade-off between them.
        This means DiskANN is not truly scalable to the scale of datasets and is not suitable for switching between multiple indices because loading all PQ vectors onto DRAM can take too long time.

        In this paper, we propose \aisaq: \aisaqdesc, a novel method of index data placement. Our implementation achieved only $\sim$10 MB of DRAM usage regardless of the scale of datasets. Our optimization for data placement in the storage index brings minor latency degradation compared to DiskANN, despite larger portion of the index are sitting in the storage. 
        Moreover, index switch between multiple billion-scale datasets can be performed in millisecond-order and makes cost advantage in multiple-server environments.
        
        The main contributions of this paper are follows:
        
        \begin{itemize}
            \item \aisaq query search requires only $\sim$10 MB of RAM for any size of dataset including SIFT1B billion-scale dataset, without changing graph topology and high recall of original DiskANN.
            \item Thanks to data placement optimization on SSD, \aisaq achieves millisecond-order latency for >95\% 1-recall@1 with tiny memory usage.
            \item \aisaq achieved negligible index load time before query search. Applications can flexibly switch between multiple billion-scale corpus corresponding to users' requests. 
            \item Index switch between different datasets in the same vector space can further reduce the switching time to sub-millisecond order, due to sharing their PQ centroid vectors.
            \item Multiple-server search for large-scale indices makes cost advantage of total DRAM and SSD consumption.
        \end{itemize}

        As a concurrent work, LM-DiskANN\cite{pan2023lm} was published before this paper's submission. However, our work proves the scalability mentioning billion-scale datasets and multiple-server systems for emerging size of datasets in the future. We also propose a new metric called index switching time, whose result reinforces the scalability for larger datasets. Both of these points are not stated in their paper.

    \section{Preliminaries}

        \subsection{Graph-Based ANNS Algorithms}

            In a dataset $X$ containing $N$ points of $d$-dimensional vector $\bm{x}_i \in \mathbb{R}^d, i = 1, \dots, N$ and a given query vector $\bm{q} \in \mathbb{R}^d$, nearest neighbor search (NNS) is a task to compute the nearest neighbor vector $ \bm{x} = \operatorname*{argmin}_{\bm{x}_i}d(\bm{x}, \bm{q})$ in a distance metric $d(\cdot, \cdot)$. Being a brute-force computation, NNS takes $O(Nd)$ time complexity and is not optimal for large-scale datasets.

            To reduce the query search time, methods to approximately compute nearest neighbor vectors are proposed, which are called approximate nearest neighbor search (ANNS).
            
            Among various approaches of ANNS, a graph-based method uses directed graph as an index. In the index construction phase, the algorithm regards the dataset vectors $\bm{x}_i$ as a node $\bm{v}_i$ and makes some edges between nodes based on their distances. In the query search phase, the algorithm starts from the entry point and move along the edges, which navigates to the candidate point of the nearest neighbor $\bm{x}$ of the query $\bm{q}$. These graph-based ANNS methods empirically acheive $O(d\log{N})$ time complexity\cite{arya1993approximate}. Some ANNS algorithms uses compressed vectors like PQ ones instead of full-precision vectors to reduce memory usage and computational cost by compromising precise distance calculation.

        \subsection{Index Switch for Multiple Sources}
            Some RAG systems retrieve information from external sources using ANNS methods. For further improvement of LLMs with RAG, ANNS methods need to ``switch'' between multiple indices of different domains corresponding to users' or applications' requests. For example, an LLM which generates latest news across multiple fields such as economy, politics and enternainment needs to generate a single answer with multiple sources of those fields. 
            Another example is an LLM chain which uses multiple sources. Represented by LangChain\cite{Chase_LangChain_2022}, it is common to use a pipeline with mutiple-time sequential LLM inferences which uses the output from an LLM inference as the input for the next inference.
            In such situations, it is possible that a RAG inference requires a different source than the one used in the preceding inference. Existing ANNS methods have to retain all of the indices on RAM or load data of an index every time from storage to switch between multiple indices. This leads to large DRAM usage or long index load time.

        \subsection{DiskANN}
            DiskANN\cite{jayaram2019diskann}, one of graph-based ANNS methods, constructs a flat directed graph index by an algorithm called Vamana. In the query search phase, we traverse the graph according to a search algorithm called beam search derived from greedy search.

            \begin{figure}
                \centering
                \begin{algorithm}[H]
                    \Description{Beam search algorithm of DiskANN updates top-$w$ candidates in every transition and sorts candidates by full-precision distances after graph transition.}
                    \caption{DiskANN Beam Search with re-ranking}
                    \label{alg:beamsearch_reranking}
                    \begin{algorithmic}
                        \STATE \textbf{Data}: Graph $G$ with entrypoint $\bm{s}$, query $\bm{q}$, number of top candidates $k$, beamwidth $w$,and search list size $L \ge k$
                        \STATE \textbf{Result}: \textit{PQ compressed} node set $\mathcal{L}$ and 
                            \textit{full-precision} node set $\mathcal{V}$, each containing top-$k$ neighbor candidates of $\bm{q}$
                        \STATE $\mathcal{L} \leftarrow \{\bm{s}\}, \mathcal{V} \leftarrow \emptyset$
     
                        \WHILE{$\mathcal{L} \setminus \mathcal{V} \ne \emptyset$}
                            \STATE let $\mathcal{P}$ denote top-$w$ closest nodes to $\bm{q}$ in $\mathcal{L}$
    
                            \FOR{$\bm{p} \in \mathcal{P}$}
                                \STATE get node chunk of $\bm{p}$ from storage
                            \ENDFOR
                                
                            \STATE append \textit{PQ vectors} of $N_{out}(\mathcal{P})$ from RAM to $\mathcal{L}$
                            \STATE append \textit{full-precision vectors} of $\mathcal{P}$ from node chunk to $\mathcal{V}$
                            \STATE $\mathcal{L} \leftarrow \mathcal{L} \cup  N_{out}(\mathcal{P}), \mathcal{V} \leftarrow \mathcal{V} \cup \mathcal{P} $
                            \IF{$|\mathcal{L}| > L$}
                                \STATE set $\mathcal{L}$ with top-$L$ closest nodes in \textit{PQ space} to $\bm{q}$ in $\mathcal{L}$ 
                            \ENDIF
                        \ENDWHILE
                        \STATE sort $\mathcal{V}$ by their \textit{full-precision} distance to $\bm{q}$
                        \STATE return top-$k$ closest nodes to $\bm{q}$ in $\mathcal{V}$  
                    \end{algorithmic}
    
                \end{algorithm}            
            \end{figure}

\begin{figure}
                \centering
                \includegraphics[width=\columnwidth]{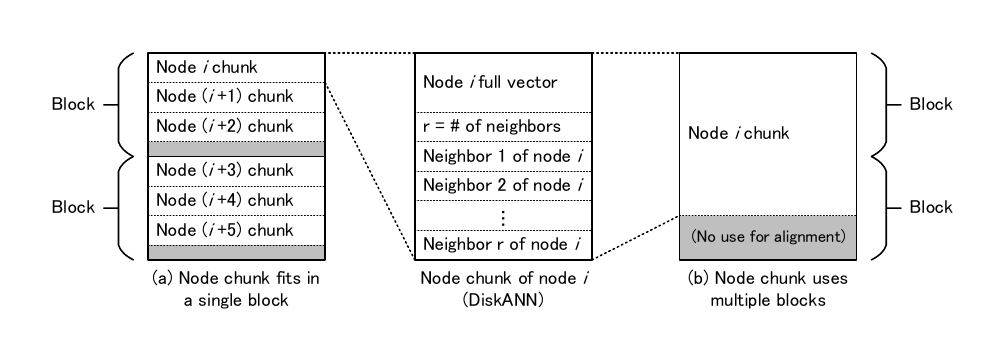}
                \Description{
                    A node chunk of DiskANN is composed of full-precision vector, number and IDs of outneighbors.
                    Node chunks are aligned in blocks.
                }
                \caption{Node chunk details and alignment in LBA blocks}
                \label{fig:node_info_diskann}
            \end{figure}

            Algorithm \ref{alg:beamsearch_reranking} shows detailed procedure of beam search used in DiskANN. The \textit{PQ compressed vector} is stored on DRAM and is used for search path determination. On the other hand, the \textit{full-precision vector} and IDs of outneighbors $N_{out}(\bm{v}_i)$ of $\bm{v}_i$ is written in the continuous LBA (Logical Block Address) space on the storage like SSD (Figure \ref{fig:node_info_diskann}). In this paper, we call this information chunk of a specific node $\bm{v}_i$ ``node chunk'' of $\bm{v}_i$. In each graph hop, DiskANN reads the node chunk of $\bm{v}_i$ and determines the next search path comparing \textit{PQ} distance between query and each of $N_{out}(\bm{v}_i)$, looking up \textit{PQ vectors} retained on the DRAM. After the graph hops finished, DiskANN sorts the candidate nodes in the search path by their \textit{full-precision} distances, which is called re-ranking.

            With $b_{full}$ bytes of each full-precision vector, max outdegree $R$, 
            and $b_{num}$ bytes (usually 4 bytes) to express a node ID or the outdegree, the size of a single node chunk is $B_{DiskANN} = b_{full} + b_{num}(R + 1)$. Operation systems dispatches I/O requests to storage devices in ``blocks'', whose size is $B = 4 \text{ KB}$ in most OS settings. In most cases, $B_{DiskANN} \le B$ stands and a block contains a single or multiple node chunk(s) in this case (Figure \ref{fig:node_info_diskann} (a)). If $B_{DiskANN} > B$, the chunk uses muitiple blocks (Figure \ref{fig:node_info_diskann} (b)). In both cases, node chunk which doesn't fit in the rest space after previous node chunk in a block is aligned to the start of the next block.
            The query search of DiskANN dispatches I/O requests to read $\left\lceil\frac{B_{DiskANN}}{B}\right\rceil$ block(s) used by a single node chunk. For example, assuming thar block size is 4 KB, the I/O size is 4 KB in Figure \ref{fig:node_info_diskann} (a) and 8 KB in Figure \ref{fig:node_info_diskann} (b).

        \subsection{Drawbacks of DiskANN}
            \textbf{Memory Usage}.
                DiskANN loads PQ vectors of all nodes in its dataset. This means memory usage of DiskANN is roughly proportional to the scale of the dataset $N$. For instance, an index constructed from SIFT1B dataset with 32 byte per PQ vector, which corresponds to $1/4$ compression of the original vector, requires 32 GB of RAM. 
                Increasing the PQ compression ratio brings recall degradation even with re-ranking, and the memory usage is still proportional to $N$.

            \textbf{Spatial Efficiency}.
                In the DiskANN beam search for a query, the number of nodes used for distance computation is much smaller than the dataset scale $N$. Even for multiple queries, few nodes are accessed frequently, while most nodes are unused and just occupy a large amount of RAM.
            
            \textbf{Index Switch Time}.
                Conventional graph-based ANNS methods including DiskANN cost time to load vector data to DRAM before the query search starts, which also increases together with $N$. To realize index switch, using muitiple large-scale datasets and switching the corpus corresponding to applications' requests, conventional methods need to keep all vector data of multiple datasets in DRAM to avoid the loading time.

    \section{Proposed Method}

    To address the above problems, we propose a novel method named \aisaq: 
    \aisaqdesc
    which offloads PQ vector data from DRAM to storage and aims almost zero memory usage with negligible latency degradation.

    \begin{figure*}
        \includegraphics[width=.8\textwidth]{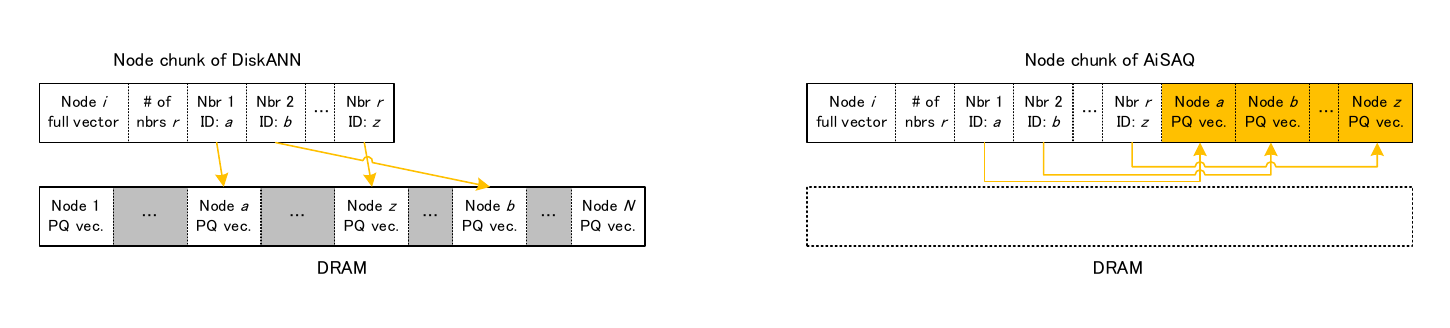}
        \Description{
            \aisaq obtains PQ vectors of outneighbors from its node chunk, while DiskANN looks up the memory.
        }
        \caption{Data placements of a node chunk and memory of DiskANN (left) and proposed method \aisaq (right)}
        \label{fig:aisaq_explanation}
    \end{figure*}

    \subsection{Methodology}
        When the DiskANN beam search is located in a node $\bm{v}(i)$, the PQ vectors of $N_{out}(\bm{v}_i)$ correspond to their IDs stored in the node chunk of $\bm{v}(i)$. This means these PQ vectors themselves can also be stored in the node chunk, which is the main idea of \aisaq (Figure \ref{fig:aisaq_explanation}).
        \aisaq also needs to adjust the node chunks which has larger size than DiskANN to fill blocks effectively and to minimize the I/O latency increment. With $b_{PQ}$ bytes of each PQ vector, the node chunk size of \aisaq is $B_{\aisaq} = B_{DiskANN} + Rb_{PQ} = b_{full} + b_{num} + R(b_{num} + b_{PQ})$. It is recommended to adjust the maximum degree $R$ so that 
        $B_{\aisaq} \le nB$ or $B_{\aisaq} \le \frac{B}{n} (n \in \mathbb{N})$ 
        stands.

        In the query search phase, PQ vectors of $N_{out}(\bm{v}_i)$ are obtained from the node chunk of $\bm{v}_i$ in the storage instead of memory and used to compute PQ distances to the query.  
        Utilizing unused area of blocks in DiskANN by filling with PQ vectors,
        this needs no additional I/O requests for getting PQ vectors and enables distance computations of each hop within the node chunk without DRAM. 
        The PQ vectors can be discarded after distance computations in each hop.
        Since the PQ vector of a specific node is obtained in the previous hop, the above algorithm can't get PQ vectors of the entrypoints. Therefore, we keep only PQ vectors of $n_{ep}$ (1 in most cases) entrypoints in DRAM. In this method, we have to keep at most $(R+n_{ep})$ PQ vectors in DRAM at once, which is independent of and much smaller than $N$, and can aim near-zero memory usage.

    \subsection{Implementation}
        We implemented \aisaq index creation algorithm based on the existing DiskANN programs \cite{diskann-github} and its query search algorithm from scratch, only reusing some utilities from DiskANN. From Vamana graph and PQ vectors generated by DiskANN, a single \aisaq index file is generated. Our implementation prioritized to lower the memory footprint, which does not support additional features of DiskANN such as filtering\cite{gollapudi2023filtered} and dynamic indexing\cite{singh2021freshdiskann}.

\section{Evaluation}
    In this section, we conducted some query search experiments using \aisaq index files and compared with DiskANN results in the same construct and search conditions.

    \subsection{Datasets and Experimental Conditions}

        \begin{table}
            \caption{Datasets and index build parameters used for \aisaq experiments}
            \label{tbl:datasets}
            \begin{minipage}{\columnwidth} 
                \centering                   
                \begin{tabular}{lrrr}
                    \toprule    
                    Dataset & SIFT1M & SIFT1B & KILT E5 22M \\
                    \midrule
                    \# of vectors $N$ & 1,000,000 & 1,000,000,000 & 22,220,792 \\
                    Dimensionality $d$ & 128 & 128 & 1024 \\ 
                    Data type & \texttt{float} & \texttt{uint8} & \texttt{float} \\ 
                    Distance metric & Euclid & Euclid & MIPS \\
                    \midrule
                    Maximum outdegree $R$ & 56 & 52 & 69 \\
                    \# of PQ subvectors $b_{PQ}$ & 128 & 32 & 128 \\
                    \bottomrule
                \end{tabular} 
            \end{minipage}
        \end{table}

        We benchmarked \aisaq and DiskANN with various scales of SIFT datasets\cite{sift2010} and KILT E5 dataset, the latter of which is a 1024-dimensional vector dataset generated by applying e5-large-v2\cite{wang2022e5} embedding to KILT\cite{petroni-etal-2021-kilt} knowledge source.

        Table \ref{tbl:datasets} shows specifications of the above datasets used  for DiskANN and \aisaq indices. Each PQ subvector of all experiments in this section is categorized into 256 clusters and can be represented in 8 bits (1 byte), which means the number of PQ subvectors can be translated into the size $b_{PQ}$ in bytes of each PQ compressed vectors. We built one-shot indices, adjusting the maximum degree $R$ for each dataset so that node chunk effectively fills a single block or multiple blocks. In each search, we fixed the beamwidth $w=4$ and changed the candidate list size $L$.

        We conducted single-server query search experiments (\ref{subsec:memory_usage} to \ref{subsec:index_switch}) on an AWS EC2 i4i.8xlarge instance which has 2x 3750 GB instance stores tied up by RAID 0.
        The environment of multiple-server experiments is detailed in \ref{subsec:cost_analysis}.

        
    \subsection{Memory Usage}\label{subsec:memory_usage}
    
        \begin{table}
            \caption{Memory Usage (MB) by DiskANN and \aisaq}
            \label{tbl:mem_usage}
            \begin{tabular}{lrrr}
                \toprule
                & $b_{PQ}$ & DiskANN & \aisaq (ours) \\
                \midrule
                SIFT1M & 128 bytes & 146 & 11 \\ 
                SIFT1B & 32 bytes & 31,303 &  11 \\
                KILT E5 22M & 128 bytes & 2,803 & 14 \\
                \bottomrule
            \end{tabular}
        \end{table}
        
        We measured peak memory usage of DiskANN and \aisaq query search by \texttt{/usr/bin/time} command. Since both of the algorithms load the entire query files onto DRAM and that portion is not essential for algorithm comparison itself, measurement of memory usage was executed only for 10 queries without groundtruth data in each dataset. Table \ref{tbl:mem_usage} shows the memory usage of DiskANN and \aisaq given the same index construction and search parameters for each dataset. While DiskANN required up to 32 GB of RAM in SIFT1B dataset, \aisaq used at most 14 MB of RAM even for billion-scale query search.

        The breakdown of \aisaq's memory usage inspected with some commands like \texttt{size} and \texttt{htop} revealed that 300 KB of memory out of the entire 11 MB is used for the search program itself and statically allocated variables, and 5.8 MB for shared libraries. Since we reuse some of DiskANN's utilities for current \aisaq implementation, better implementation of \aisaq beam search may further reduce memory usage.

    \subsection{Query Search Time}

        \begin{figure*}
            \begin{minipage}[t]{0.7\textwidth}%
                \centering
                \begin{minipage}[t]{0.3\linewidth}%
                    \includegraphics[width=\linewidth]{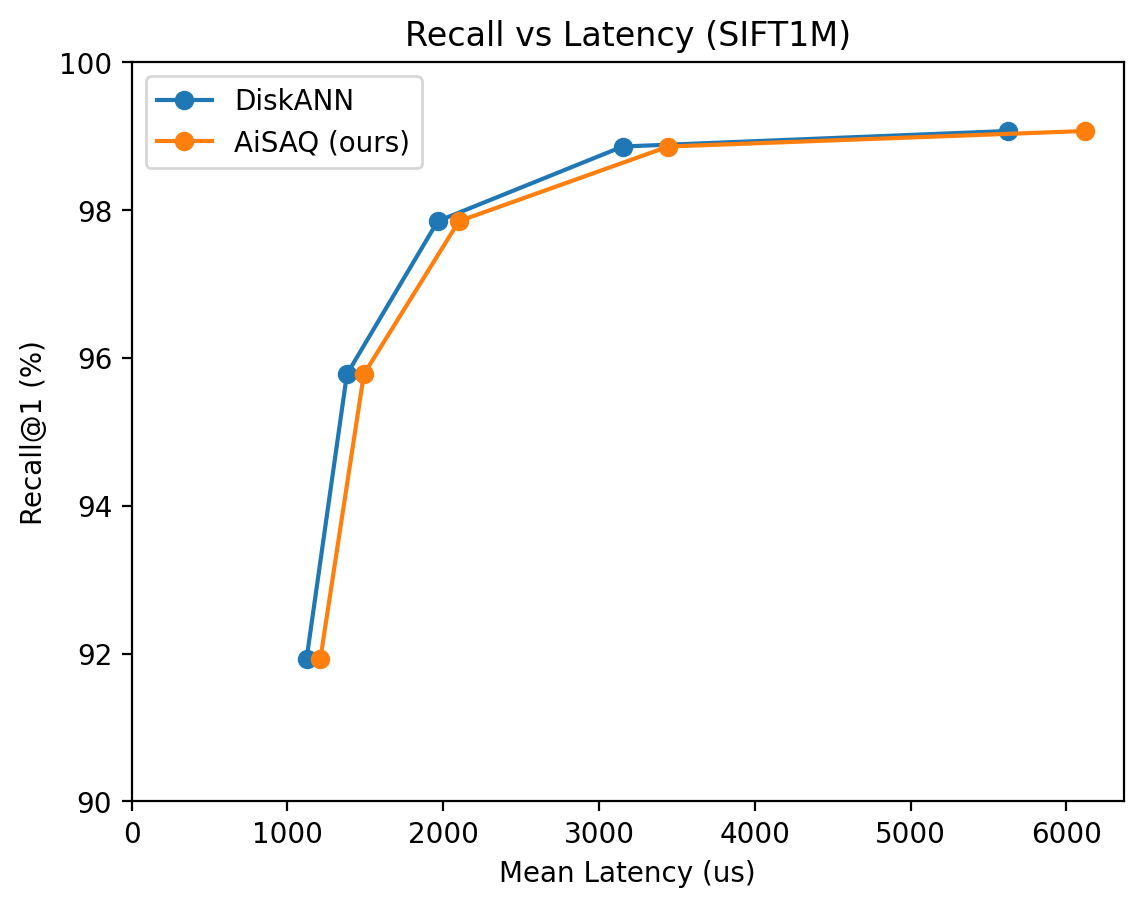}%
                    \subcaption{SIFT1M}                    
                \end{minipage}
                \begin{minipage}[t]{0.3\linewidth}%
                    \includegraphics[width=\linewidth]{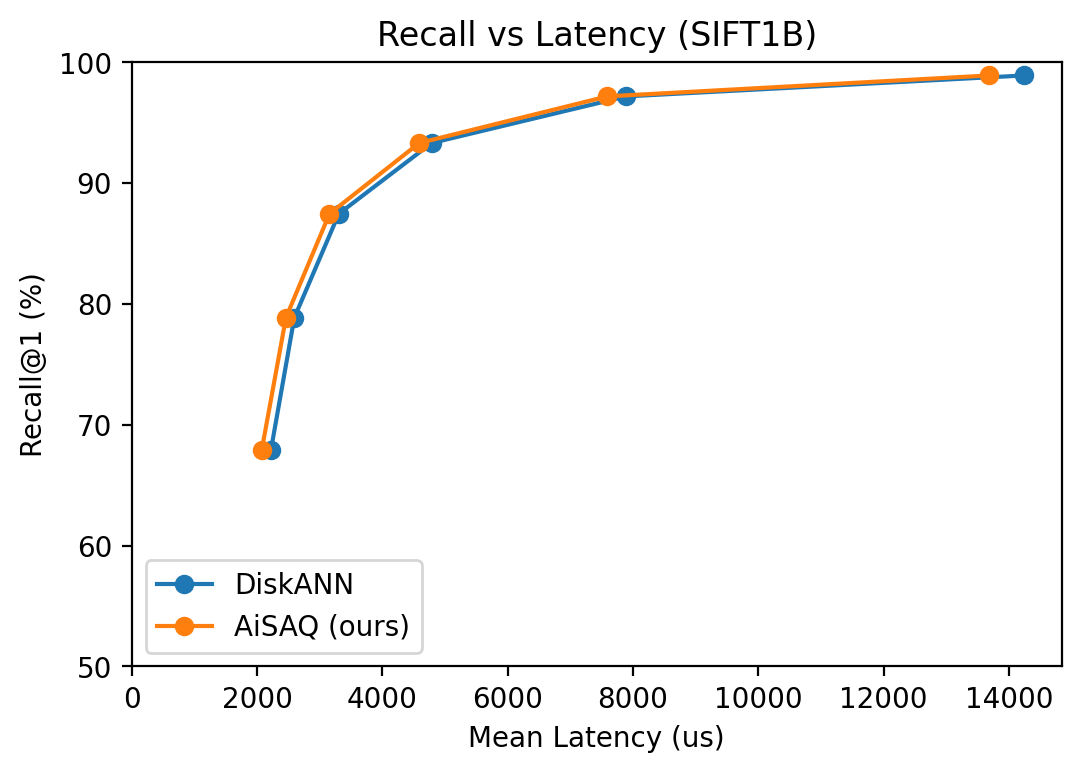}%
                    \subcaption{SIFT1B}
                \end{minipage}
                \begin{minipage}[t]{0.3\linewidth}%
                    \includegraphics[width=\linewidth]{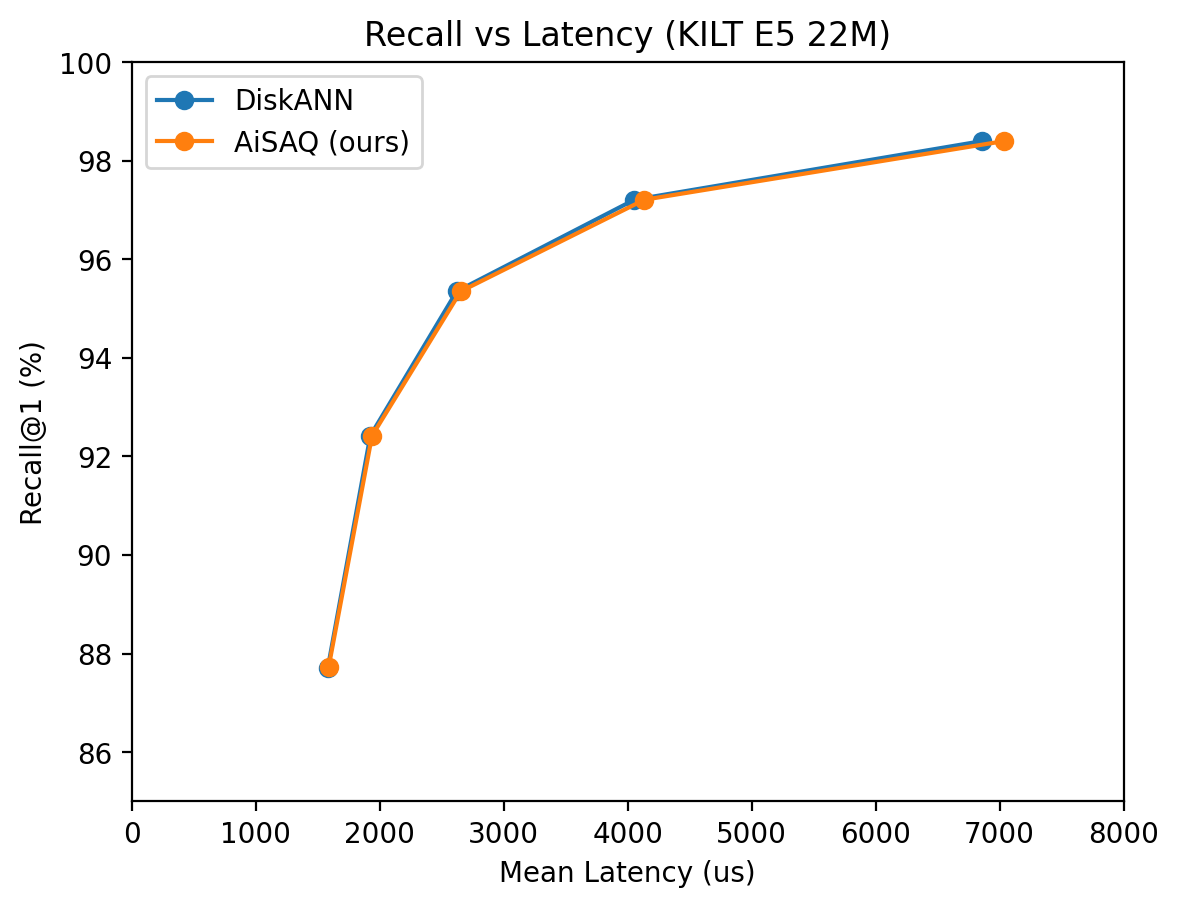}%
                    \subcaption{KILT E5 22M}
                \end{minipage}
                \Description{
                    \aisaq's query search latencies outperforms DiskANN, while recall@1 is identical.
                }
                \caption{Recall vs latency graphs of DiskANN and \aisaq}
                \label{fig:recall_latency_comparison}    
            \end{minipage}
            \begin{minipage}[t]{0.25\textwidth}%
                \centering
                \includegraphics[width=\linewidth]{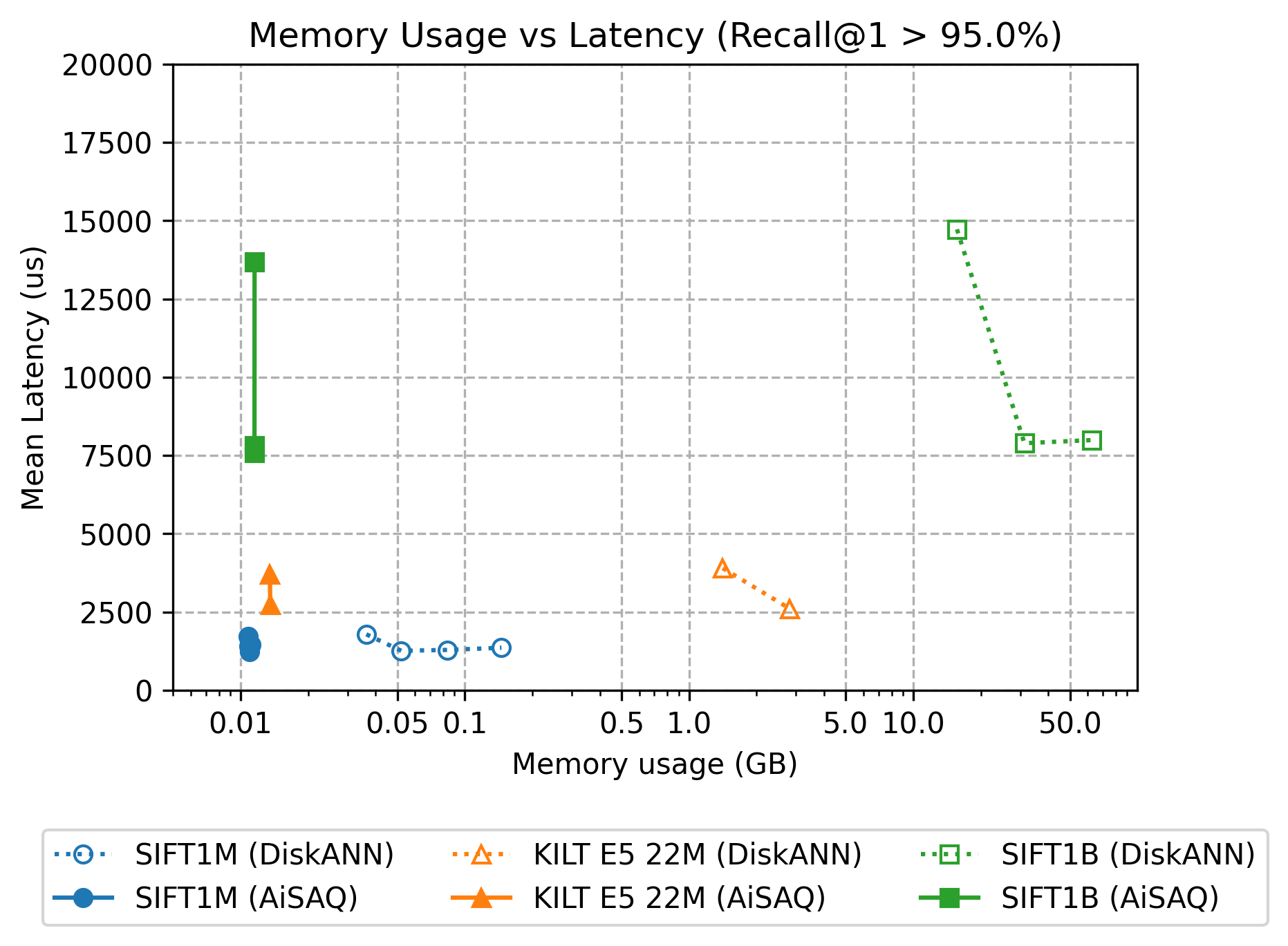}
                \Description{
                    In the high-recall area, \aisaq keeps $\sim$11 MB memory usage for all datasets,
                    while DiskANN suffers lower QPS to keep recall@1 when memory usage is reduced.
                }
                \caption{Latency vs memory usage of DiskANN and \aisaq}
                \label{fig:qps_mem_usage_comparison}    
            \end{minipage}
        \end{figure*}

        Figure \ref{fig:recall_latency_comparison} shows recall-vs-latency plots of DiskANN and \aisaq in each dataset. Since \aisaq does not change the graph topology itself, recall@1 is identical to DiskANN in the same search condition. In some datasets like SIFT1M and KILT E5 22M, \aisaq needs more blocks for a node chunk than DiskANN and the I/O request size for query search also becomes larger. Such datasets lead to I/O latency degradation compared with the same search conditions of DiskANN. 
        Thanks to the I/O queueing system of SSDs and our optimization with shorter time for distance computation in CPU, however, the latency degradation is not critical.
        On the other hand, \aisaq performs faster on SIFT1B because the I/O request sizes are the same $B_{\aisaq} = B_{DiskANN} = \SI{4}{KB}$.
        
        We also conducted query search experiments for indices with various PQ vector sizes $b_{PQ}$ to change the memory usage of DiskANN search. Figure \ref{fig:qps_mem_usage_comparison} shows search latencies of DiskANN and \aisaq search to achieve > 95\% recall@1 of each $b_{PQ}$ by memory usage. DiskANN searches configured to consume less RAM require slower latencies with large search list size $L$ to ensure high recall@1. In contrast to DiskANN's memory-latency tradeoff, \aisaq keeps the same small memory usage regardless of PQ vector size $b_{PQ}$, which means \aisaq can keep both of near-zero memory usage and millisecond-order search latencies in the high-recall area. 
        
    \subsection{Index Switch}\label{subsec:index_switch}
        \begin{table}
            \caption{Index load time (ms) of DiskANN and \aisaq}
            \label{tbl:index_load_time}
            \begin{tabular}{lrr}
                \toprule
                & DiskANN & \aisaq (ours) \\
                \midrule
                SIFT1M & 46.8 & 0.6 \\ 
                SIFT1B & 16,437.4 & 0.6 \\ 
                KILT E5 22M & 1,121.4 & 2.0 \\ 
                \bottomrule
            \end{tabular}
        \end{table}

        Table \ref{tbl:index_load_time} shows the load time of DiskANN and \aisaq indices before query search starts. While DiskANN's load time increases with the dataset scale, \aisaq keeps constantly short time since there is no need to load PQ vectors of the entire datasets. The load time of \aisaq indices itself is the same order as its query search latencies. Higher-dimensional PQ centroids of KILT E5 22M brings longer time than the other datasets.
        
        \begin{table}
            \caption{KILT E5 2.2M subset index switch time (ms) of \aisaq with and without PQ centroids reloading}
            \label{tbl:index_switch_time}
                \begin{tabular}{lrr}
                    \toprule
                        & Index switch time (ms) \\
                    \midrule
                    DiskANN & 119.2 \\
                    \aisaq (PQ centroids reloading) & 1.9 \\ 
                    \aisaq (shared PQ centroids) & 0.3 \\
                    \bottomrule
                \end{tabular}
        \end{table}

        In applications of LLMs with RAG, it is possible that multiple datasets share the vector space (for example, images and documents are encoded to the same space by each encoder). In such situations, datasets are also able to share their PQ centroids. To obtain multiple datasets which can share the centroids, we devided KILT E5 22M dataset into 10 subsets, employing the PQ centroids generated by the whole 22M dataset. Then we built an index from each subset and searched across these indices with or without reloading the PQ centroids file. From table \ref{tbl:index_switch_time}, sharing the PQ centroids, which only needs to load 4 KB metadata for an index, significantly reduced index switch time even compared to \aisaq with reloading. This method is especially effective for high-dimensional datasets like KILT E5.
    
    \subsection{Cost Analysis and Multiple-Server System}\label{subsec:cost_analysis}
        \begin{figure}
            \centering
            \includegraphics[width=\columnwidth]{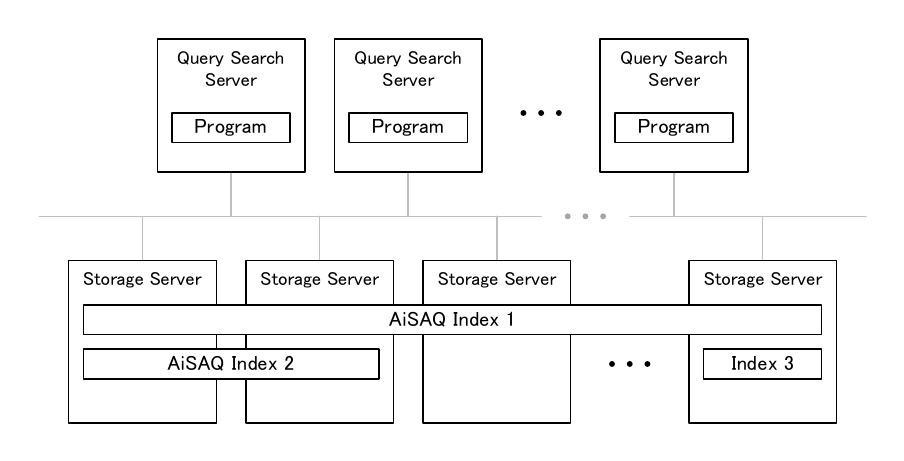}
            \caption{Concept of multiple-server \aisaq system}
            \label{fig:multi_server_system}
        \end{figure}
        \begin{figure}
            \centering
            \includegraphics[width=0.8\columnwidth]{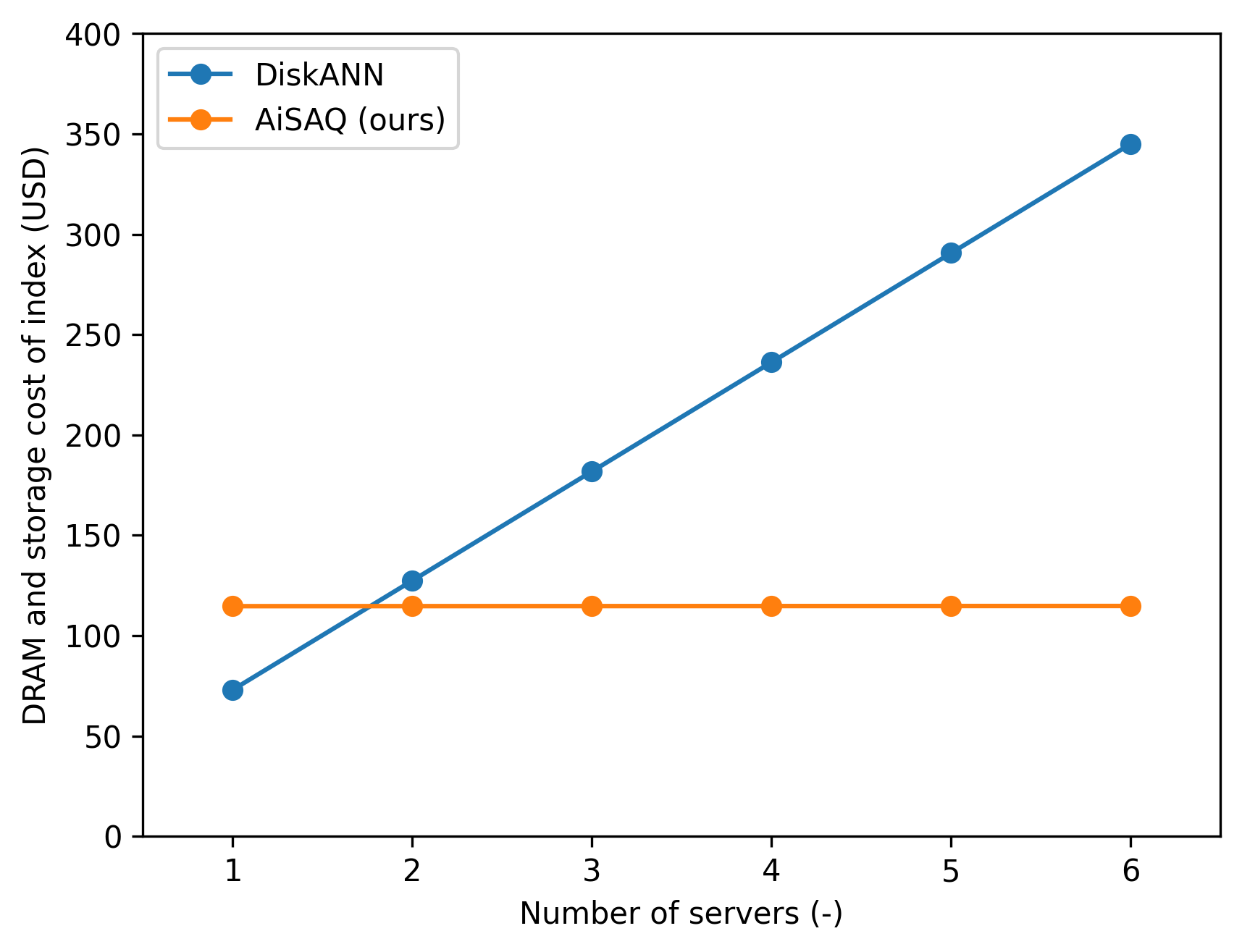}
            \caption{Cost estimation of DiskANN and \aisaq (SIFT1B)}
            \label{fig:cost_estimation}
        \end{figure}
        \begin{table}
            \caption{DiskANN and \aisaq comparison with 6 containers search (SIFT1B)}
            \label{tbl:multi_server_comparison}
            \begin{tabular}{lrr}
                \toprule
                    & DiskANN & \aisaq \\
                \midrule
                    Total memory usage (MB) & 183,405 & 66 \\
                    Average index load time (ms) & 14319.25 & 1.7 \\
                    Estimated DRAM \& SSD cost (USD) & 344 & 103 \\
                \bottomrule
            \end{tabular}
        \end{table}
        \aisaq achieves similar high performance to DiskANN with extremely small memory usage and index load time, which results in inflated SSD index size. To retain the PQ vector information, an \aisaq index requires $RNb_{PQ}$ bytes of storage, while DiskANN requires $Nb_{PQ}$ bytes of DRAM. Maximum outdegree of the graph $R$ is typically set around 60, while the cost per bit of DRAM is approximately 30 times of that of SSD (DRAM: 1.8 USD/GB vs SSD: 0.054 USD/GB, obtained from DRAMeXchange.com\cite{dramexchange}). This means \aisaq is not more cost effective than DiskANN in a single server.

        However, it is possible that query search scales out with multiple servers like Figure \ref{fig:multi_server_system} to handle increased search requests and larger indices. Such situations makes cost advantages to \aisaq. Figure \ref{fig:cost_estimation} shows estimated DRAM and SSD costs of DiskANN and \aisaq. DiskANN search using $n$ servers requires $n$ times of DRAM capacity, while \aisaq takes only the same capacity of storage as the case with a single server. \aisaq overtakes DiskANN in terms of the resource cost in more than 2 search servers, which is reasonable for multiple-server search for larger-scale distributed ANNS system.

        Table \ref{tbl:multi_server_comparison} shows experimental results and cost estimation of DiskANN and \aisaq on multiple-server system. We used Lustre distributed filesystem with 3 storage servers for the index storage. \aisaq search programs are executed on 6 Docker containers each regarded as a server on a single physical machine with an EPYC 7402P CPU and 224 GB of DRAM. All machines are connected by RDMA over 200 Gb ethernet. DiskANN with 6 containers consumes much more DRAM than \aisaq, which can be the main factor of total cost. \aisaq resulted in slower index load time than experiment with direct-attached SSD because of the slower latency over ethernet, but still reasonablly fast time, while DiskANN did not acheive practical load time.

    \section{Conclusion and Future Works}
        In this paper, we proposed \aisaq, a novel method of ANNS data offloading to storage. Our method achieved billion-scale ANNS with only $\sim$10 MB memory usage without significant degradation of performance. \aisaq will be applicable to all graph-based ANNS algorithms, 
        and multiple-server system with lower costs.
        
        In addition, reducing the index load time before query search enabled index switch between multiple large-scale dataset for various users' requests. That will enable LLMs with RAG to employ more simple index addition or filter search algorithms.

    \bibliography{references}
    \bibliographystyle{acm}

\end{document}